\documentclass[aip,jcp,reprint,a4paper]{revtex4-1}
\usepackage{graphicx,xcolor,amsmath,amssymb,amsfonts,physics}
\usepackage[utf8]{inputenc}
\usepackage{longtable}

\usepackage[version=4]{mhchem} 

\usepackage[colorlinks = true,
            linkcolor = blue,
            urlcolor  = blue,
            citecolor = red,
            anchorcolor = blue]{hyperref}

\begin{document}

\title{TREXIO: A File Format and Library for Quantum Chemistry}

\newcommand{\LCPQ}{Laboratoire de Chimie et Physique
Quantiques (LCPQ), Universit\'e de Toulouse (UPS) and CNRS,
Toulouse, France}
\newcommand{\UTwente}{MESA+ Institute for Nanotechnology, University of Twente, P.O. Box 217, 7500 AE Enschede, The Netherlands}
\newcommand{\UVSQ}{Université Paris-Saclay, UVSQ, LI-PaRAD}
\newcommand{\MaxPlanck}{Max Planck Institute for Solid State Research, Heisenbergstrasse 1, 70569 Stuttgart, Germany}
\newcommand{\SISSA}{International School for Advanced Studies (SISSA), Via Bonomea 265, 34136, Trieste, Italy}
\newcommand{\Qubit}{Qubit Pharmaceuticals, Incubateur Paris Biotech Santé, 24 Rue du Faubourg Saint Jacques, 75014 Paris, France}

\author{Evgeny Posenitskiy}
\affiliation{\LCPQ{}}
\affiliation{\Qubit{}}

\author{Vijay Gopal Chilkuri}
\affiliation{\LCPQ{}}
\affiliation{Aix Marseille Univ, CNRS, Centrale Marseille, ISM2, Marseille, France}

\author{Abdallah Ammar}
\affiliation{\LCPQ{}}

\author{Michał Hapka}
\affiliation{Faculty of Chemistry, University of Warsaw, ul. L. Pasteura 1, 02-093 Warsaw, Poland}
\author{Katarzyna Pernal}
\affiliation{Institute of Physics, Lodz University of Technology, ul. Wolczanska 217/221, 93-005 Lodz, Poland}

\author{Ravindra Shinde}
\author{Edgar Josué Landinez Borda}
\author{Claudia Filippi}
\affiliation{\UTwente{}}

\author{Kosuke Nakano}
\affiliation{Research and Services Division of Materials Data and Integrated System, National Institute for Materials Science (NIMS), Tsukuba, Ibaraki 305-0047, Japan}
\affiliation{\SISSA{}}

\author{Otto Kohulák}
\affiliation{\SISSA{}}
\affiliation{\LCPQ{}}
\author{Sandro Sorella}
\affiliation{\SISSA{}}

\author{Pablo de Oliveira Castro}
\author{William Jalby}
\affiliation{\UVSQ}

\author{Pablo López Ríos}
\author{Ali Alavi}
\affiliation{\MaxPlanck}

\author{Anthony Scemama}
\email{scemama@irsamc.ups-tlse.fr}
\affiliation{\LCPQ{}}

\newcommand{\cf}[1]{#1}
\newcommand{\cc}[1]{#1}




\begin{abstract}
\begin{minipage}[t]{0.75\linewidth}
TREXIO is an open-source file format and library developed for the storage and manipulation of data produced by quantum chemistry calculations.
It is designed with the goal of providing a reliable and efficient method of storing and exchanging wave function parameters and matrix elements, making it an important tool for researchers in the field of quantum chemistry.
In this work, we present an overview of the TREXIO file format and library.
The library consists of a front-end implemented in the C programming language and two different back-ends: a text back-end and a binary back-end utilizing the HDF5 library which enables fast read and write operations. 
It is compatible with a variety of platforms and has interfaces for the Fortran, Python, and OCaml programming languages. In addition, a suite of tools has been developed to facilitate the use of the TREXIO format and library, including converters for popular quantum chemistry codes and utilities for validating and manipulating data stored in TREXIO files. 
The simplicity, versatility, and ease of use of TREXIO make it a valuable resource for researchers working with quantum chemistry data.
\end{minipage}
\vtop{
 \vskip-1ex
  \hbox{\phantom{..}
    \includegraphics[width=0.15\linewidth]{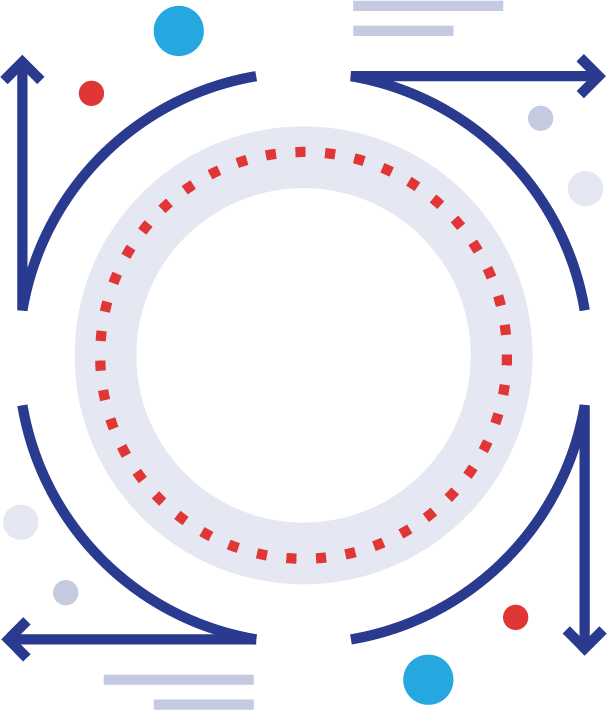}
  }
}
\end{abstract}

\maketitle

\section{Introduction}

Quantum chemistry relies on quantum mechanics to explain and predict the properties and behaviors of atoms, molecules, and materials.
Although density functional theory (DFT) is one of the most widely used approaches thanks to its excellent ratio between computational cost and accuracy, another important tool is wave function theory (WFT), which describes the behavior of a quantum system in terms of its wave function.
In order to perform WFT calculations, it is necessary to manipulate a large number of parameters, such as the expansion coefficients of the wave function and the matrix elements of the Hamiltonian operator.
These parameters are typically numerous and difficult to handle, making it important to have a robust and efficient method for storing and accessing them.

Reproducible research remains a challenging topic, despite recent advances such as the introduction of the FAIR (findable, accessible, interoperable, reusable) data principles.\cite{wilkinson_2016}
A key aspect of reproducibility is software interoperability, which refers to the ability of different programs to work together and exchange information, allowing different systems to communicate and exchange data in order to function as a cohesive whole.
Interoperable software is prevalent nowadays and is a key component of the Unix philosophy.\cite{mcilroy_1978}
In Unix shells, the most straightforward application of software interoperability is made through the use of the \emph{pipe} operator, where the output of a program is the input of another program.
Similarly, shell scripts are created through the composition of smaller programs, exchanging data through files or pipes.

A major challenge of reproducible research is the uniformity of input/output (I/O) data within a particular research domain.
The Unix philosophy recommends the use of text files because they are architecture-independent, readable in any language, and can be read as a stream, which is useful for making programs communicate over a network.
However, storing data in a text format can result in large file sizes and conversion from ASCII to binary format can be computationally expensive for large data sets.
To address this concern, domain-specific binary formats have been developed, such as the Joint Photographic Experts Group (JPEG) format\cite{jpeg} for digital images and the Moving Picture Experts Group (MPEG) format\cite{mpeg} for videos.
These binary formats are utilized through standardized application programming interfaces (API).

In the field of wave function theory such a standard format and API is still lacking, and the purpose of the TREXIO library presented in this article is to fill this gap.
This paper is organized as follows: firstly, a brief overview of the related work is presented.
Secondly, the TREXIO format for the electronic wave functions is introduced together with some details concerning the internal representation and the associated API.
Finally, some applications are demonstrated with a major focus on the interoperability achieved within the TREX Center of Excellence in Exascale Computing~\cite{TREXwebsite} due to the use of the TREXIO format.

\section{Related work}

It is worth mentioning that there have been several efforts to unify the data formats within different subdomains of quantum chemistry.
Probably one of the earliest works in this direction was the definition of the Crystallographic Information File (CIF) for establishing databases of crystal structures.\cite{hall_1991}
A few years later, the Chemical Markup Language (CML)\cite{murray-rust_1999,murray-rust_2011} was introduced.
It is a format based on the Extensible Markup Language (XML) which is used to describe chemical data: molecules, chemical properties, reactions, spectra, materials, \textit{etc}.
With formats like CIF or CML, the burden of following a standard is placed on the code \emph{writing} the data. 
As a consequence, any tool that can read the format will be able to interpret the data without needing to understand the specific code that was used to produce it.
This means that data can be easily shared and reused across different programs,
and new tools can be developed to work with the format without needing to know anything about the code used to produce the data.

Recently, the \texttt{cclib} Python package \cite{oboyle_2008}, originally developed for performing computational chemistry calculations, has accumulated several internal converters capable of parsing and transforming the output of different programs into the internal representation called \texttt{ccData}.
A similar approach has been taken by the developers of IOData \cite{verstraelen_2021}, who have implemented converters and parsers for commonly used programs and their output files.
However, there is currently no unified data representation or API that can be integrated into quantum chemistry codes to improve interoperability.
Consequently, each time a given program modifies its input/output formatting, the IOData package must be adapted accordingly and promptly, which poses an additional challenge for maintainers.
More recently, consolidated efforts have given rise to QCSchema \cite{smith_2021}, which provides an API-like access to data generated by existing quantum chemistry codes, thereby addressing the issue of dependence on the output file's formatting style.
In this case, the responsibility for adhering to conventions falls on the code \emph{reading} the data, as it must be aware of the conventions chosen by the code that generated the data.
With the Electronic Structure Common Data Format (ESCDF) \cite{oliveira_2020} and its associated library, codes that write data can supply metadata to assist codes that read data in comprehending the organization of the data in the files.
Hence, ESCDF aims to provide low-level tools and flexibility to facilitate the exchange of large datasets between codes with high-performance I/O.
While this greatly reduces the difficulty of understanding conventions for developers reading the data, they may still need to apply different conversions depending on the code that generated the data.
Consequently, implementing support for ESCDF may require more effort on the part of code developers compared to using a standardized format such as CML.

\cf{
Another popular format for storing quantum chemistry data is the Gaussian\cite{gaussian} \texttt{fchk} format. While it is a proprietary format specific to the Gaussian software package, its compatibility with several other software programs has contributed to its extensive utilization.
However, the format's proprietary and closed-source nature prevents external developers from 
improving the format, leaving enhancements and compatibility updates solely in the hands of Gaussian developers.
}

Recently, the \texttt{mwfn}\cite{lu_2022} format was introduced with the primary goal of enhancing the existing solutions such as \texttt{wfn},\cite{gaussian} \texttt{wfx},\cite{keith_2014} and Molden\cite{schaftenaar_2000} formats, which were designed to store parameters of molecular orbitals and atomic basis sets in view of reconstructing the one-particle density matrix.
Although \texttt{mwfn} is an improvement on these other formats, it does not allow the user to store enough information for a wave function coming from a configuration interaction (CI) or coupled cluster (CC) calculation.

For post-Hartree-Fock calculations, the \texttt{FCIDUMP} format\cite{knowles_1989} has become a \textit{de facto} standard because of its simplicity.
It is a text-based format that only contains minimal information for building the second-quantized Hamiltonian, namely the one- and two-electron integrals in the basis of molecular orbitals (MO), the number of electrons and information about the spin state and orbital symmetries. The nuclear coordinates and basis set are not saved in \texttt{FCIDUMP} files.
The text format makes its adoption extremely simple, but it has a very high impact on the performance since \texttt{FCIDUMP} files are usually large.
Although very practical, the use of the \texttt{FCIDUMP} format has other important limitations than efficiency.
Once a program has computed a post-Hartree-Fock wave function using an \texttt{FCIDUMP} file as an input, the parameters of the basis set and the molecular orbitals may have been lost unless they were stored in a separate file in another format. 
Although configuration interaction or coupled cluster calculations can be performed using \texttt{FCIDUMP} files, this format is too limited to be used for quantum Monte Carlo (QMC) calculations, which require \emph{all} the wave function parameters.

The Q5Cost\cite{borini_2007,scemama_2008,rossi_2014} initiative was one of the first attempts aiming at standardizing the WFT data by introducing both a format and the API to interact with it.
With Q5Cost, it was possible to store all the wave function parameters of CI expansions together with the basis set, molecular orbitals, and even electron repulsion integrals.
The Q5Cost library was relying on the Hierarchical Data Format version 5 (HDF5)\cite{hdf5} to provide efficient I/O and keep the data well organized in the file.
Nevertheless, Q5Cost had some severe drawbacks.
First, Q5Cost was written in Fortran which made its use tedious in other programming languages such as C\texttt{++} or Python.
In addition, to be able to interpret a Q5Cost file, it was often necessary to know which code had generated it.
Indeed, most WFT codes have different conventions in terms of normalization of the basis functions, ordering of the atomic orbitals, \textit{etc}, and no conversion into a unique internal representation was imposed by the library. So the burden of understanding conventions was still on the shoulders of the readers of the files.
Finally, Q5Cost had important technical limitations: the Q5Cost library was intended to be used as a compiled Fortran module (a so-called \texttt{.mod} file), that depended on the compiled Fortran modules provided by the HDF5 library.
As the format of the compiled Fortran modules is specific to the compiler vendor and even to the version of the compiler, the Q5Cost library could not be simply linked as an external library to any code.
Using the Q5Cost library in a Fortran code imposed that the user's code was compiled with the same Fortran compiler as the one that was used to compile both the HDF5 Fortran modules and the Q5Cost library.
This contamination of dependencies could lead to some important impact on the performance of the user's code, and the only solution to solve that problem was to compile many different versions of the HDF5 Fortran interface and Q5Cost library with multiple compilers and compiler versions.

The TREXIO initiative, heavily influenced by the Q5Cost project, aims to propose a standard format and library for wave function calculations. This initiative seeks to leverage the strengths of the Q5Cost project and learn from its design flaws that hindered its widespread adoption.
One of the key improvements we aim to achieve is to shift the effort of adopting a format and conventions to the side of the code writing the data. This way, the files will be easily readable without any prior knowledge by any code, similar to CML or JPEG.

\section{The TREXIO format}

The TREXIO format (version 2.3.0) is designed to store all the necessary information to represent a wave function, including: the number of up- and down-spin electrons, nuclear coordinates and charges, basis set and effective core potential (ECP) parameters, atomic and molecular orbital parameters, Slater determinants and CI coefficients, configuration state function (CSF) definitions, and metadata related to the description of excited states.
It is also capable of storing data required for the computation of the wave function, such as one- and two-electron integrals, numerical integration grids used in DFT calculations, and one- and two-particle reduced density matrices.

One notable feature of TREXIO is that it is self-contained, meaning that all the parameters needed to recreate the wave function are explicitly stored within the file, eliminating the need for external databases.
For example, instead of storing the name of a basis set (such as cc-pVDZ), the actual basis set parameters used in the calculation are stored.
All data are stored in atomic units for simplicity.

The data in TREXIO are organized into \emph{groups}, each containing multiple \emph{attributes} defined by their \emph{type} and \emph{dimensions}.
Each attribute within a group corresponds to a single scalar or array variable in a code.
In what follows, the notation \texttt{<group>.<attribute>} will be used to identify an attribute within a group.
For example, \texttt{nucleus.charge} refers to the \texttt{charge} attribute in the \texttt{nucleus} group.
It is an array of type \texttt{float} with dimension \texttt{nucleus.num}, the attribute describing the number of nuclei.
For simplicity, the singular form is always used for the names of groups and attributes.

\subsection{Data types}

So that TREXIO can be used in any language, we use a limited number of data types.
\cf{
It is important to keep in mind that these types are abstract in the sense that they are defined independently of their implementation, and are not tied to any specific representation on a computer.}
The main data types are \texttt{int} for integers, \texttt{float} for floating-point values, and \texttt{str} for character strings.
\cf{The real and imaginary parts of complex numbers are stored separately as \texttt{float}s.}
To minimize the risk of integer overflow and accuracy loss, numerical data types are stored using 64-bit representations by default.
However, in specific cases where integers are bounded (such as orbital indices in four-index integrals), the smallest possible representation is used to reduce the file size.
The API presented in the next section handles any necessary type conversions.

There are also two types derived from \texttt{int}: \texttt{dim} and \texttt{index}.
\texttt{dim} is used for dimensioning variables, which are positive integers used to specify the dimensions of an array.
In the previous example, \texttt{nucleus.num} is a dimensioning variable that specifies the dimensions of the \texttt{nucleus.charge} array.
\texttt{index} is used for integers that correspond to array indices, because some languages (such as C or Python) use zero-based indexing, while others (such as Fortran) use one-based indexing \cf{by default}.
For convenience, values of the \texttt{index} type are shifted by one when TREXIO is used in one-based languages to be consistent with the semantics of the language.  

Arrays can be stored in either dense or sparse formats.
If the sparse format is selected, the data is stored in coordinate format.
For example, the element \texttt{A(i,j,k,l)} is stored as a quadruplet of integers $(i,j,k,l)$ along with the corresponding value.
Typically, \cf{one- and} two-dimensional arrays are stored as dense arrays, while arrays with higher dimensions are stored in sparse format.

\subsection{Stored data}

In this section, we provide a comprehensive overview of the data that can be stored in TREXIO files.
A complete list of the groups and attributes is available \cf{as supplementary information} or in the documentation of the library.
In both resources, multi-dimensional arrays are expressed in column-major order, meaning that elements of the same column are stored contiguously.

\subsubsection{Metadata}

In order to facilitate the archiving of TREXIO files in open-data repositories, users have the option to store metadata in the \texttt{metadata} group.
This includes the names of the codes that were used to create the file, a list of authors, and a textual description.
This allows for more information about the file to be easily accessible and transparent.

\subsubsection{System information}

The chemical system consists of nuclei and electrons, where the nuclei are
considered as fixed point charges with Cartesian coordinates.
The wave function is stored in the spin-free formalism,\cite{pauncz_1989} and therefore, it is necessary
to explicitly store the number of spin-up ($N_\uparrow$) and spin-down ($N_\downarrow$) electrons.
These numbers correspond to the normalization of the spin-up and spin-down single-particle reduced density matrices.

Certain calculations, such as DFT calculations, require the use of a numerical integration grid. 
The \texttt{grid} group provides information for storing grids, inspired by the data required by the \texttt{numgrid} software.\cite{bast_2021,burkardt_2010}

To keep things simple, TREXIO can only store a single wave function per file.
When working with excited states, it is often the case that multiple states only differ in their CI coefficients, while other parameters (such as geometry, basis set, molecular orbitals, etc.) are the same.
To facilitate the storage of multiple states, TREXIO provides the option to store all the data needed to describe one state in a main file, along with the names of additional TREXIO files that contain only the state-specific parameters.

\subsubsection{Basis set}

In the \texttt{basis} group, the atomic basis set is defined as a list of shells.
\cf{Each shell $i$ is centered at a center $A_i$, has a specified angular momentum $l_i$, and a radial function $R_i$.
The radial function is a linear combination of $N_{\text{prim}\,i}$ \emph{primitive} functions, which can be Slater type orbitals (STO, $p=1$) or Gaussian type orbitals (GTO, $p=2$).
These primitive functions are parameterized by exponents $\gamma_{ki}$ and coefficients $a_{ki}$:
}
\begin{equation}
    R_i(\mathbf{r}) = \mathcal{N}_i \vert\mathbf{r}-\mathbf{R}_{A_i}\vert^{n_i}
    \sum_{k=1}^{N_{\text{prim}\,i}} a_{ki}\, f_{ki}(\gamma_{ki},p)\,
    e^{ -\gamma_{ki} \vert \mathbf{r}-\mathbf{R}_{A_i} \vert ^p} .
    \label{eq:Rs}
\end{equation}
Different codes have different normalization practices, so it is necessary to store normalization factors in the TREXIO file to ensure that it is self-contained and does not rely on the client program having the ability to compute overlap integrals.
Some codes assume that the contraction coefficients are applied to \emph{normalized} linear combinations of primitives, so a normalization constant $f_{ki}$ for each primitive must also be stored.
Some codes assume that the functions $R_i$ are normalized, requiring the computation of an additional normalization factor, $\mathcal{N}_i$.

\subsubsection{Atomic orbitals}

The \texttt{ao} group in TREXIO contains information related to the expansion of the shells in the basis set into atomic orbitals (AOs).
For example, a $p$-shell is expanded into three AOs: $p_x$, $p_y$, and $p_z$.
AOs are defined as follows:
\begin{equation}
   \chi_i (\mathbf{r}) = \mathcal{N}_i'\, P_{\eta(i)}(\mathbf{r})\, R_{s(i)} (\mathbf{r})
\end{equation}
where $i$ is the atomic orbital index, $P$ refers to either polynomials or spherical harmonics, and $s(i)$ specifies the shell on which the AO is expanded.

$\eta(i)$ denotes the chosen angular function.
The AOs can be expressed using real spherical harmonics or polynomials in Cartesian coordinates.
In the case of real spherical harmonics, the AOs are ordered as $0, +1, -1, +2, -2, \dots, +m, -m$.
In the case of polynomials, the canonical (or alphabetical) ordering is used, 
\begin{eqnarray}
p & : & p_x, p_y, p_z \nonumber \\
d & : & d_{xx}, d_{xy}, d_{xz}, d_{yy}, d_{yz}, d_{zz} \nonumber \\
f & : & f_{xxx}, f_{xxy}, f_{xxz}, f_{xyy}, f_{xyz}, f_{xzz}, f_{yyy}, f_{yyz}, f_{yzz}, f_{zzz} \nonumber \\
\vdots \nonumber
\end{eqnarray}
Note that for \(p\) orbitals in \cf{real spherical harmonics}, the ordering is $0,+1,-1$ which corresponds to $p_z, p_x, p_y$.

$\mathcal{N}_i'$ is a normalization factor that allows for different normalization coefficients within a single shell, as in the GAMESS\cite{barca_2020} convention where each individual function is unit-normalized. 
Using GAMESS convention, the normalization factor of the shell $\mathcal{N}_d$ (Eq.~\ref{eq:Rs}) in the \texttt{basis} group is appropriate for instance for the $d_z^2$ function (i.e. $\mathcal{N}_{d}\equiv\mathcal{N}_{z^2}$) but not for the $d_{xy}$ AO, so the correction factor $\mathcal{N}_i'$ for $d_{xy}$ in the \texttt{ao} groups is the ratio $\frac{\mathcal{N}_{xy}}{\mathcal{N}_{z^2}}$.

\subsubsection{Effective core potentials}

An effective core potential (ECP) $V_A^{\text{ECP}}$ can be used to replace the core electrons of atom A. It can be expressed as:\cite{trail_2017} 
\cf{
\begin{equation}
   V_A^{\text{ECP}} =
   V_{A \ell_{\max}+1} +
   \sum_{\ell=0}^{\ell_{\max}}
   \delta V_{A \ell}\sum_{m=-\ell}^{\ell} | Y_{\ell m} \rangle \langle Y_{\ell m} |
\end{equation}
The first term in this equation is attributed to the local channel, while the remaining terms correspond to non-local channel projections.
$\ell_{\max}$ refers to the maximum angular momentum in the non-local component of the ECP.
The functions \(\delta V_{A \ell}\) and \(V_{A \ell_{\max}+1}\) are parameterized as:
\begin{eqnarray}
   \delta V_{A \ell}(\mathbf{r}) &=&
   \sum_{q=1}^{N_{q \ell}}
   \beta_{A q \ell}\, |\mathbf{r}-\mathbf{R}_{A}|^{n_{A q \ell}}\,
   e^{-\alpha_{A q \ell} |\mathbf{r}-\mathbf{R}_{A}|^2 } \nonumber\\
   V_{A \ell_{\max}+1}(\mathbf{r}) &=& -\frac{Z_\text{eff}}{|\mathbf{r}-\mathbf{R}_{A}|}+\delta V_{A \ell_{\max}+1}(\mathbf{r})
\end{eqnarray}
where $Z_\text{eff}$ is the effective nuclear charge of the center.
}
All the parameters can be stored in the \texttt{ecp} group.

\subsubsection{Molecular orbitals}

The \texttt{mo} group is devoted to the storage of the molecular orbitals (MOs).
MO coefficients are stored in a two-dimensional array, with additional information such as symmetries or occupation numbers stored in separate arrays.
It is also possible to store the spin to enable the description of unrestricted Hartree-Fock or unrestricted Kohn-Sham determinants.

\subsubsection{Hamiltonian matrix elements}

One-electron integrals can be stored in the AO and MO bases in the groups \texttt{ao\_1e\_int} and \texttt{mo\_1e\_int}, respectively.
Similarly, two-electron integrals can be stored in the AO and MO bases in the groups \texttt{ao\_2e\_int} and \texttt{mo\_2e\_int}, respectively.
One-electron integrals are stored as two-dimensional arrays, while two-electron integrals are stored in a sparse format, with a quadruplet of indices and the corresponding value stored for each non-zero integral.
The order of the indices follows Dirac's bra-ket notation.

It is also possible to store a low-rank representation of the two-electron integrals, obtained via a Cholesky decomposition.

\subsubsection{CI expansion}

The wave function $\Psi$ can be represented as a combination of Slater determinants $D_I$:
\begin{equation}
   \ket{\Psi} = \sum_I C_I \ket{D_I}
\end{equation}
In the \texttt{determinant} group of a TREXIO file, the definition of these Slater determinants, as well as the configuration interaction (CI) expansion coefficients, can be stored.
Each Slater determinants is represented as a Waller-Hartree double determinant,\cite{pauncz_1984} i.e. the product of a determinant with $\uparrow$-spin electrons and a determinant with $\downarrow$-spin electrons.
To enable the storage of arbitrary CI expansions and to reduce the storage size, the determinants are stored as pairs of \cf{binary} strings: one for the $\uparrow$ spin sector and one for the $\downarrow$ spin.
\cf{Each binary string has a length equal to the number of MOs, with the $i$-th bit set to one if and only if the $i$-th MO is included in the determinant.}
As the creation of these \cf{binary} strings may be tedious, we provide some helper
functions to transform lists of orbital indices into \cf{binary} strings.
If the orbital indices are not in increasing order, a reordering is made and the user is informed if a change of sign is needed in the corresponding CI coefficient.

Alternatively, the wave function may be expanded in a basis of configuration state functions (CSFs),
\begin{equation}
   \ket{\Psi} = \sum_I \tilde{C}_I \ket{\psi_I}.
\end{equation}
where each CSF $\psi_I$ is a linear combination of Slater determinants.
The \texttt{csf} group allows for the storage of the CSF expansion coefficients, as well as the matrix $\langle D_I | \psi_J \rangle$ in a sparse format.
This enables the projection of the CSFs onto the basis of Slater determinants.

\subsubsection{Amplitudes}

The wave function may also be expressed in terms of the action of the
cluster operator \(\hat{T}\):
\begin{equation}
   \hat{T} = \hat{T}_1 + \hat{T}_2 + \hat{T}_3 + \dots
\end{equation}
on a reference wave function \(\Psi\), where \(\hat{T}_1\) is the
single excitation operator,
\begin{equation}
   \hat{T}_1 = \sum_{ia} t_{i}^{a}\, \hat{a}^\dagger_a \hat{a}_i,
\end{equation}
\(\hat{T}_2\) is the double excitation operator,
\begin{equation}
   \hat{T}_2 = \frac{1}{4} \sum_{ijab} t_{ij}^{ab}\, \hat{a}^\dagger_a \hat{a}^\dagger_b \hat{a}_j \hat{a}_i,
\end{equation}
\textit{etc}. Indices $i$, $j$, $a$ and $b$ denote molecular orbital indices.

Wave functions obtained with perturbation theory or configuration
interaction are of the form:
\begin{equation}
  |\Phi\rangle = \hat{T}|\Psi\rangle
\end{equation}
and coupled-cluster wave functions are of the form:
\begin{equation}
  |\Phi\rangle = e^{\hat{T}}| \Psi \rangle
\end{equation}

The reference wave function $\Psi$ is stored using the \texttt{determinant} and/or \texttt{csf} groups, and the amplitudes are stored using the \texttt{amplitude} group.
The attributes with the \texttt{exp} suffix correspond to exponentialized operators.

\subsubsection{Reduced density matrices}

The reduced density matrices, stored in the \texttt{rdm} group, are defined in the basis of molecular orbitals.

The $\uparrow$-spin and $\downarrow$-spin components of the one-body density matrix are given by
\begin{eqnarray}
  \gamma_{ij}^{\uparrow}   &=& \langle \Psi | \hat{a}^{\dagger}_{j\alpha}\, \hat{a}_{i\alpha} | \Psi \rangle \\
  \gamma_{ij}^{\downarrow} &=& \langle \Psi | \hat{a}^{\dagger}_{j\beta} \, \hat{a}_{i\beta}  | \Psi \rangle
\end{eqnarray}
and the spin-summed two-body density matrix is
\begin{equation}
   \gamma_{ij} = \gamma^{\uparrow}_{ij} + \gamma^{\downarrow}_{ij}
\end{equation}

The $\uparrow \uparrow$,  $\downarrow \downarrow$, and $\uparrow \downarrow$ components of the two-body density matrix are given by
\begin{eqnarray}
  \Gamma_{ijkl}^{\uparrow \uparrow} &=&
     \langle \Psi | \hat{a}^{\dagger}_{k\alpha}\, \hat{a}^{\dagger}_{l\alpha} \hat{a}_{j\alpha}\, \hat{a}_{i\alpha} | \Psi \rangle \\
  \Gamma_{ijkl}^{\downarrow \downarrow} &=&
     \langle \Psi | \hat{a}^{\dagger}_{k\beta}\, \hat{a}^{\dagger}_{l\beta} \hat{a}_{j\beta}\, \hat{a}_{i\beta} | \Psi \rangle \\
  \Gamma_{ijkl}^{\uparrow \downarrow} &=&
     \langle \Psi | \hat{a}^{\dagger}_{k\alpha}\, \hat{a}^{\dagger}_{l\beta} \hat{a}_{j\beta}\, \hat{a}_{i\alpha} | \Psi \rangle + \nonumber \\ 
&&     \langle \Psi | \hat{a}^{\dagger}_{l\alpha}\, \hat{a}^{\dagger}_{k\beta} \hat{a}_{i\beta}\, \hat{a}_{j\alpha} | \Psi \rangle,
\end{eqnarray}
and the spin-summed one-body density matrix is
\begin{equation}
   \Gamma_{ijkl} = \Gamma_{ijkl}^{\uparrow \uparrow} +
     \Gamma_{ijkl}^{\downarrow \downarrow} + \Gamma_{ijkl}^{\uparrow \downarrow}.
\end{equation}

\subsubsection{Correlation factors}

Explicit correlation factors can be introduced in the wave function, such as in QMC, $F_{12}$, or transcorrelated methods.

In the current version of the library, it is possible to store two different types of Jastrow factors.
The Jastrow factor is an $N$-electron function which multiplies the reference wave function expansion: $\Psi = \Phi \times \exp(J)$, where
\begin{equation}
   J(\mathbf{r},\mathbf{R}) = J_{\text{eN}}(\mathbf{r},\mathbf{R}) + J_{\text{ee}}(\mathbf{r}) + J_{\text{eeN}}(\mathbf{r},\mathbf{R}).
\end{equation}
In the following, we use the notations $r_{ij} = |\mathbf{r}_i - \mathbf{r}_j|$ and $R_{i\alpha} = |\mathbf{r}_i - \mathbf{R}_\alpha|$, where indices $i$ and $j$ correspond to electrons and $\alpha$ to nuclei.

The first form of Jastrow factor is the one used in the CHAMP\cite{champ} program.\cite{guclu_2005}
$J_{\text{eN}}$ contains electron-nucleus terms:
\begin{align}
   J_{\text{eN}}(\mathbf{r},\mathbf{R}) =  \sum_{i=1}^{N_\text{elec}} \sum_{\alpha=1}^{N_\text{nucl}} \Bigg[ &
   \frac{a_{1,\alpha}\, f_\alpha(R_{i\alpha})}{1+a_{2,\alpha}\,
   f_\alpha(R_{i\alpha})} \nonumber \\
 & + \sum_{p=2}^{N_\text{ord}^a} a_{p+1,\alpha}\, [f_\alpha(R_{i\alpha})]^p - J_{\text{eN}}^\infty
   \Bigg]
\end{align}
$J_{\text{ee}}$ contains electron-electron terms:
\begin{align}
   J_{\text{ee}}(\mathbf{r}) =
   \sum_{i=1}^{N_\text{elec}} \sum_{j=1}^{i-1}
   \Bigg[ &
   \frac{\frac{1}{2}\big(1 + \delta^{\uparrow\downarrow}_{ij}\big)\,b_1\, f_{\text{ee}}(r_{ij})}{1+b_2\, f_{\text{ee}}(r_{ij})} \nonumber \\
 & +
   \sum_{p=2}^{N_\text{ord}^b} b_{p+1}\, [f_{\text{ee}}(r_{ij})]^p  - J_{\text{ee},ij}^\infty
   \Bigg]
\end{align}
 where $\delta^{\uparrow\downarrow}_{ij}$ is zero when the electrons $i$ and $j$ have the same spin, and one otherwise.
$J_{\text{eeN}}$ contains electron-electron-nucleus terms:
\begin{align}
   J_{\text{eeN}}(\mathbf{r},\mathbf{R}) = 
    \sum_{\alpha=1}^{N_{\text{nucl}}}
     \sum_{i=1}^{N_{\text{elec}}}
      \sum_{j=1}^{i-1}
       \sum_{p=2}^{N_{\text{ord}}}
        \sum_{k=0}^{p-1}
         \sum_{l=0}^{p-k-2\delta_{k,0}} 
           c_{lkp\alpha} \left[ g_{\text{ee}}({r}_{ij}) \right]^k \nonumber \\
            \left[ \left[ g_\alpha({R}_{i\alpha}) \right]^l + \left[ g_\alpha({R}_{j\alpha}) \right]^l \right]
             \left[ g_\alpha({R}_{i\,\alpha}) \,
  g_\alpha({R}_{j\alpha}) \right]^{(p-k-l)/2}, 
\end{align}
$c_{lkp\alpha}$ being non-zero only when $p-k-l$ is even.
The terms $J_{\text{ee}}^\infty$ and $J_{\text{eN}}^\infty$ are shifts to ensure that $J_{\text{ee}}$ and $J_{\text{eN}}$ have an asymptotic value of zero.
$f$ and $g$ are scaling functions defined as 
\begin{equation}
   f_\alpha(r) = \frac{1-e^{-\kappa_\alpha\, r}}{\kappa_\alpha} \text{ and }
   g_\alpha(r) = e^{-\kappa_\alpha\, r},
\end{equation}
and the possible presence of an index $\alpha$ indicates that the scaling coefficient $\kappa$ depends on the atom $\alpha$.

The second form of Jastrow factor is the $\mu$ Jastrow factor\cite{giner_2021} 
\begin{equation}
   J_{\text{ee}}(\mathbf{r}) = 
   \sum_{i=1}^{N_\text{elec}} \sum_{j=1}^{i-1} r_{ij}
   \left( 1 - \text{erf}(\mu\, r_{ij})\right) - \frac{1}{\mu\sqrt{\pi}}
   e^{-(\mu\,r_{ij})^2}.
\end{equation}
It is a single parameter correlation factor that has been recently introduced in the
context of transcorrelated methods. It imposes the electron-electron cusp and is
built such that the leading order in $1/r_{12}$ of the effective
two-electron potential reproduces the long-range interaction of the
range-separated density functional theory.
An envelope function has then been introduced to cancel out the Jastrow
effects between two electrons when at least one electron is close to a nucleus,
and standard one-body terms were also introduced to avoid the expansion
of the one-body density.

As there exist multiple forms of Jastrow factors in the literature, contributions to extend this section are welcome.

\subsubsection{QMC data}

We also provide in the \texttt{qmc} group some information specific to QMC calculations.
In QMC methods, the wave function is evaluated at points in the $3N$-dimensional space, where $N$ is the number of electrons.
It might be convenient to store the coordinates of points together with the wave function, and to store the value of the wave function and the local energy $\hat{H} \Psi(\mathbf{r}) / \Psi(\mathbf{r})$ evaluated at these points, for example, to check that different codes give the same values.

\section{The TREXIO library}

The TREXIO library is written in the C language, and is licensed under the open-source 3-clause BSD license to allow for use in all types of quantum chemistry software, whether commercial or not.

The design of the library is divided into two main sections: the front-end and the back-end.
The front-end serves as the interface between users and the library, while the back-end acts as the interface between the library and the physical storage.

\subsection{The front-end}

\begin{figure}
\includegraphics[width=\linewidth]{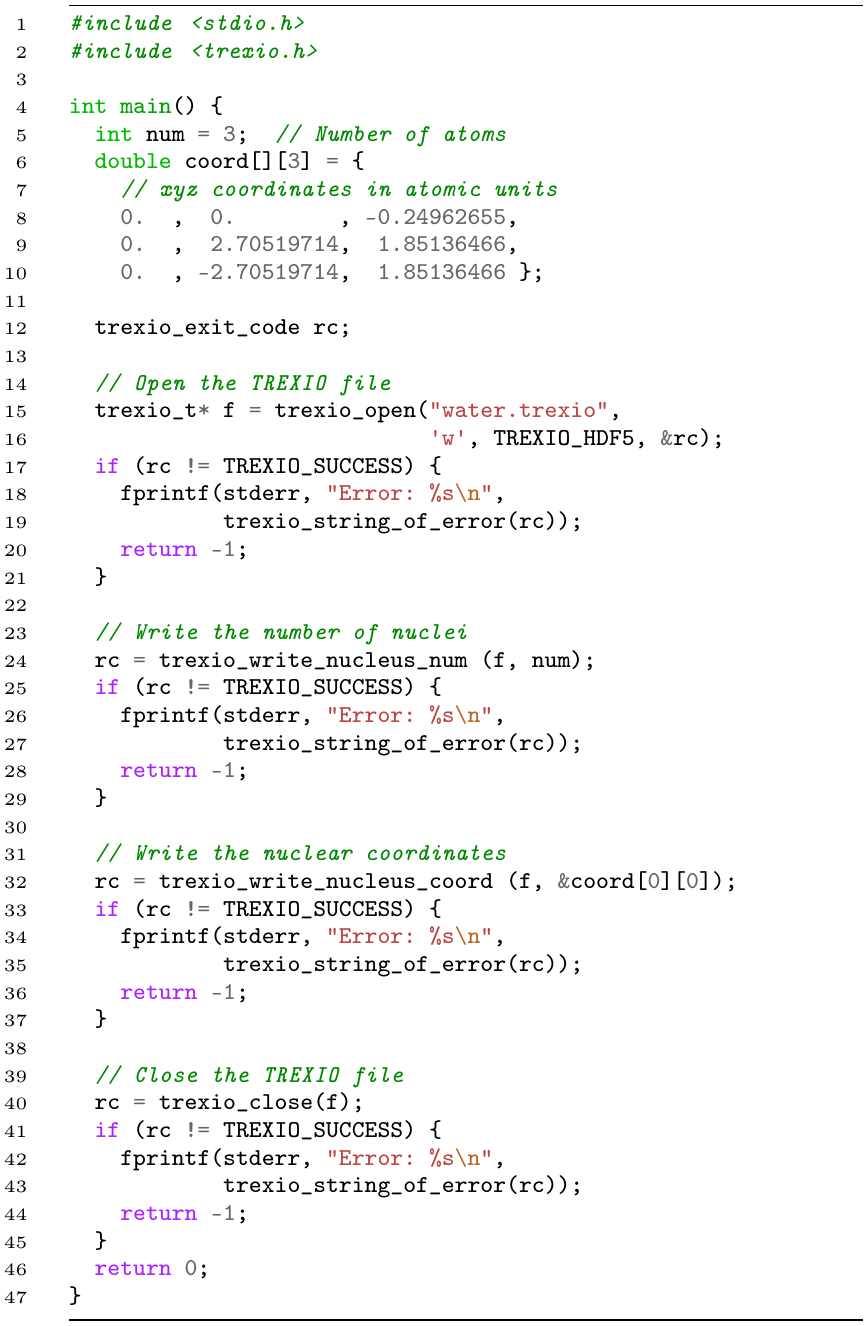}
\caption{\label{fig:example_c} C code writing the nuclear coordinates of a water molecule in a TREXIO file, with error handling.}
\end{figure}

\begin{figure}
\includegraphics[width=\linewidth]{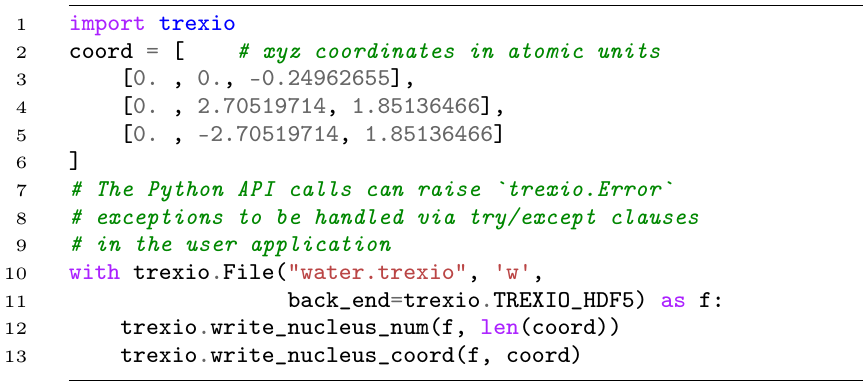}
\caption{\label{fig:example_p} Python code writing the nuclear coordinates of a water molecule in a TREXIO file.}
\end{figure}

By using the TREXIO library, users can store and extract data in a consistent and organized manner.
The library provides a user-friendly API, including functions for reading, writing, and checking for the existence of data.
The functions follow the pattern \texttt{trexio\_[has|read|write]\_<group>\_<attribute>}, where the group and attribute specify the particular data being accessed.
It also includes an error handling mechanism, in which each function call returns an exit code of type \texttt{trexio\_exit\_code}, explaining the type of error.
This can be used to catch exceptions and improve debugging in the upstream user application.
Figures~\ref{fig:example_c} and \ref{fig:example_p} show examples of usage of the TREXIO library in C and Python, respectively.

To ensure the consistency of the data, the attributes can only be written if all the other attributes on which they explicitly depend have been written.
For example, as the \texttt{nucleus.coord} array is dimensioned by the number of nuclei
\texttt{nucleus.num}, the \texttt{nucleus.coord} attribute can only be written after \texttt{nucleus.num}.
However, the library is not aware of non-explicit dependencies, such as the relation between the electron repulsion integrals (ERIs) and MO coefficients.
A complete control of the consistency of the data is therefore impossible, so the attributes were chosen to be by default \emph{immutable}.
By only allowing data to be written only once, the risk of modifying data in a way that creates inconsistencies is reduced.
For example, if the ERIs have already been written, it would be inconsistent to later modify the MO coefficients.
To allow for flexibility, the library also allows for the use of an \emph{unsafe} mode, in which data can be overwritten.
However, this mode carries the risk of producing inconsistent files, and the \texttt{metadata} group's \texttt{unsafe} attribute is set to \texttt{1} to indicate that the file has potentially been modified in a dangerous way.
This attribute can be manually reset to \texttt{0} if the user is confident that the modifications made are safe.

\subsection{The back-end}

At present, TREXIO supports two back-ends: one relying only on the C standard library to produce plain text files (the so-called text back-end), and one relying on the HDF5 library.

With the text back-end, the TREXIO ``file'' is a directory containing multiple text files, one for each group. 
This back end is intended to be used in development environments, as it gives access to the user to standard tools such as \texttt{diff} and \texttt{grep}.
In addition, text files are better adapted than binary files for version control systems such as Git, so this format can be also used for storing reference data for unit tests.

HDF5 is a binary file format and library for storing and managing large amounts of data in a hierarchical structure.
It allows users to manipulate data in a way similar to how files and directories are manipulated within the file system.
The HDF5 library provides optimal performance through its memory mapping mechanism and supports advanced features such as serial and parallel I/O, chunking, and compression filters.
However, HDF5 files are in binary format, which requires additional tools such as \texttt{h5dump} to view them in a human-readable format.
HDF5 is widely used in scientific and engineering applications, and is known for its high performance and ability to handle large data sets efficiently.

The TREXIO HDF5 back-end is the recommended choice for production environments, as it provides high I/O performance.
Furthermore, all data is stored in a single file, making it especially suitable for parallel file systems like Lustre.
These file systems are optimized for large, sequential I/O operations and are not well-suited for small, random I/O operations.
When multiple small files are used, the file system may become overwhelmed with metadata operations like creating, deleting, or modifying files, which can adversely affect performance.

In a benchmarking program designed to compare the two back-ends of the library, the HDF5 back-end was found to be significantly faster than the text back-end. 
The program wrote a wave function made up of 100 million Slater determinants and measured the time taken to write the Slater determinants and CI coefficients. 
The HDF5 back-end achieved a speed of $10.4\times10^6$ Slater determinants per second and a data transfer rate of 406 MB/s, while the text back-end had a speed of $1.1\times10^6$ determinants per second and a transfer rate of 69 MB/s. 
These results were obtained on a DELL 960 GB mix-use solid-state drive (SSD). The HDF5 back-end was able to achieve a performance level close to the peak performance of the SSD, while the text back-end's performance was limited by the speed of the CPU for performing binary to ASCII conversions.

In addition to the HDF5 and text back-ends, it is also possible to introduce new back-ends to the library.
For example, a back-end could be created to support object storage systems, such as those used in cloud-based applications\cite{liu_2018} or for archiving in open data repositories.
\cf{To use a new back-end, only a minor modification is required in the code using TREXIO: the correct back-end argument needs to be passed to the \texttt{trexio\_open} function (see Figures~\ref{fig:example_c} and~\ref{fig:example_p}).}

\subsection{Supported languages}

One of the main benefits of using C as the interface for a library is that it is easy to use from other programming languages.
Many programming languages, such as Python or Julia, provide built-in support for calling C functions, which means that it is relatively straightforward to write a wrapper that allows a library written in C to be called from another language.
In general, libraries with a C interface are the easiest to use from other programming languages, because C is widely supported and has a simple, stable application binary interface (ABI).
Other languages, such as Fortran and C\texttt{++}, may have more complex ABIs and may require more work to interface with them.

TREXIO has been employed in codes developed in various programming languages, including C, C\texttt{++}, Fortran, Python, OCaml, and Julia.
While Julia is designed to enable the use of C functions without the need for additional manual interfacing, the TREXIO C header file was automatically integrated into Julia programs using the \texttt{CBindings.jl} package.\cite{Rutkowski2023}
In contrast, specific bindings have been provided for Fortran, Python, and OCaml to simplify the user experience.

In particular, the binding for Fortran is not distributed as multiple compiled Fortran module files (\texttt{.mod}), but instead as a single Fortran source file (\texttt{.F90}).
The distribution of the source file instead of the compiled module has multiple benefits.
It ensures that the TREXIO module is always compiled with the same compiler as the client code, avoiding the compatibility problem of \texttt{.mod} files between different compiler versions and vendors.
The single-file model requires very few changes in the build system of the user's codes, and it facilitates the search for the interface of a particular function.
In addition, advanced text editors can parse the TREXIO interface to propose interactive auto-completion of the TREXIO function names to the developers.

Finally, the Python module, partly generated with SWIG\cite{beazley_1996} and fully compatible with NumPy,\cite{harris2020array} allows Python users to interact with the library in a more intuitive and user-friendly way.
Using the Python interface is likely the easiest way to begin using TREXIO and understanding its features.
In order to help users get started with TREXIO and understand its functionality, tutorials in Jupyter notebooks are available on GitHub (\url{https://github.com/TREX-CoE/trexio-tutorials}), and can be executed via the Binder platform.


\subsection{Source code generation and documentation}

Source code generation is a valuable technique that can significantly improve the efficiency and consistency of software development.
By using templates to generate code automatically, developers can avoid manual coding and reduce the risk of errors or inconsistencies.
This approach is particularly useful when a large number of functions follow similar patterns, as in the case of the TREXIO library, where functions are named according to the pattern \texttt{trexio\_[has|read|write]\_<group>\_<attribute>}.
By generating these functions from the format specification using templates, the developers can ensure that the resulting code follows a consistent structure and is free from errors or inconsistencies.

The description of the format is written in a text file in the Org format.\cite{schulte_2012}
Org is a structured plain text format, containing information expressed in a lightweight markup language similar to the popular Markdown language.\cite{leonard_2016}
While Org was introduced as a mode of the GNU Emacs text editor, its basic functionalities have been implemented in most text editors such as Vim, Atom or VS Code.

There are multiple benefits in using the Org format.
The first benefit is that the Org syntax is easy to learn and allows for the insertion of equations in \LaTeX{} syntax.
Additionally, Org files can be easily converted to HyperText Markup Language (HTML) or Portable Document Format (PDF) for generating documentation.
The second benefit is that GNU Emacs is a programmable text editor and code blocks in Org files can be executed interactively, similar to Jupyter notebooks.
These code blocks can also manipulate data defined in tables and this feature is used to automatically transform tables describing groups and attributes in the documentation into a JavaScript Object Notation (JSON) file.\cite{bray_2017,pezoa_2016}
This JSON file is then used by a Python script to generate the needed functions in C language, as well as header files and some files required for the Fortran, Python, and OCaml interfaces.

With this approach, contributions to the development of the TREXIO library can be made simply by adding new tables to the Org file, which can be submitted as \emph{pull requests} on the project's GitHub repository (\url{https://github.com/trex-coe/trexio}).
Overall, this process allows for a more efficient and consistent development process and enables contributions from a wider range of individuals, regardless of their programming skills.


\subsection{Availability and reliability}

The TREXIO library is designed to be portable and easy to install on a wide range of systems.
It follows the C99 standard to ensure compatibility with older systems, and can be configured with either the GNU Autotools or the CMake build systems.
The only external dependency is the HDF5 library, which is widely available on HPC platforms and as packages on major Linux distributions.
Note that it is possible to disable the HDF5 back-end at configuration time, allowing TREXIO to operate only with the text back-end and have zero external dependencies.
This can be useful for users who may not be able to install HDF5 on certain systems.

TREXIO is distributed as a tarball containing the source code, generated code, documentation, and Fortran interface.
It is also available as a binary \texttt{.deb} package for Debian-based Linux distributions and as packages for Guix\cite{courtes_2013}, Spack\cite{gamblin_2015} and Conda.\cite{conda_forge}
The Python module can be found in the PyPI repository, the OCaml binding is available in the official OPAM repository, and the \texttt{.deb} packages are already available in Ubuntu 23.04.

\cf{
To ensure the reliability and quality of the TREXIO library, we have 
adopted standard continuous integration and deployment practices.
For example, we use unit tests that are executed automatically using GitHub
actions whenever modifications are made to the codebase.
These tests cover a wide range of functionalities and help to identify any potential issues or bugs in the code.
Additionally, the TREXIO library is regularly used by the authors of the present paper, and as such, it is continuously tested and validated in the context of ongoing research activities.
}

\cf{TREXIO was built, tested and installed successfully on 20 different architectures supported by the Debian build farm.
Furthermore, we ensure that the quality of our code meets the requirements of the CERT coding standards,\cite{seacord_2014} and we use the \texttt{cppcheck}\cite{cppcheck} tool to validate the quality of our code.
These measures demonstrate our commitment to ensuring that the TREXIO library is a reliable and trustworthy tool.
}

\subsection{Open-Source Governance and Sustainability Strategies}

\cf{
Our approach to the development and governance of the TREXIO library follows the standard design of open-source projects, which typically involve a collaborative effort from a community of contributors.
The TREX European Center of Excellence initiated the project and proposed the first functional version of the software.
However, we consider this to be just the starting point for a larger community effort.
}

\cf{
As an open-source project, we encourage contributions from anyone interested in the development of the library.
This includes not only contributions to the codebase but also contributions to the documentation, testing, and other aspects of the project.
We believe that this collaborative approach is the key to the success of any open-source project.
}

\cf{
Regarding governance, we have a small group of maintainers who oversee the development of the project, review and merge contributions, and ensure the quality of the code.
However, we strive to make the development process as transparent and open as possible, and we encourage contributions from anyone interested in the project.
}

\cf{
Overall, our strategy for the governance and development of the TREXIO library follows the standard design of open-source projects, which emphasizes collaboration and transparency.
We believe that this approach, combined with our commitment to seeking and securing funding for the continued development and maintenance of TREXIO, will ensure the long-term success and usefulness of the library to the quantum chemistry community.
}

%

\section{Examples of applications}

The open-source Python package \texttt{trexio\_tools}\cite{trexio_tools} has been created to enhance the use of the TREXIO library and corresponding data format.
It includes converters for transforming output files from codes such as Gaussian, GAMESS,\cite{barca_2020} or PySCF\cite{sun_2020} into TREXIO files.
However, in the future, it would be preferable if the developers of these codes were to offer the option to export data in TREXIO format in order to maintain numerical precision and ensure consistency in the stored data.
In addition, the package includes utilities to convert certain data blocks from TREXIO files into FCIDUMP or Molden formats.
It also has a feature that validates the consistency of a wave function by numerically calculating overlap integrals on a grid and comparing them to the overlap matrix stored in the file.
This helps to confirm that all basis set parameters are consistent with the conventions of the original program. 

TREXIO is currently used to exchange wave function parameters between the selected CI code Quantum Package\cite{garniron_2019} and the QMC code CHAMP.\cite{champ} The QMC codes QMC=Chem\cite{scemama_2013} and TurboRVB\cite{nakano_2020} are also able to read TREXIO files, allowing for comparison of the three QMC codes using the same wave function.
TREXIO is also used to transfer integrals between Quantum Package and the FCIQMC code NECI,\cite{guther_2020} and to read density matrices produced by Quantum Package in GammCor\cite{gammcor} for symmetry-adapted perturbation theory (SAPT)\cite{jeziorski_1994} molecular interaction calculations with near-full CI density matrices.\cite{hapka_2021}
In addition, the recent development of a code for calculating electron repulsion integrals using Slater-type orbitals\cite{caffarel_2019} now produces TREXIO files, enabling FCIQMC calculations using Slater-type orbitals with NECI and similar selected CI calculations with Quantum Package, which can then be used as trial wave functions for QMC calculations.

\section{Conclusion}

The TREXIO format and library offer a convenient and flexible way to store and exchange quantum chemistry data.
Its open-source nature allows for easy integration into various software applications and its compatibility with multiple programming languages makes it accessible to a wide range of users.
The use of the HDF5 library as the default back-end ensures efficient storage and retrieval of data, while the option to disable HDF5 and use the text back-end allows for zero external dependencies.
The development of TREXIO has been driven by the need to facilitate collaboration and reproducibility in quantum chemistry research, and its adoption in various codes and projects is a testament to its usefulness in achieving these goals.
\cf{We would like to emphasize that the TREXIO library is a work in
progress, and we are committed to expanding its scope and functionality in 
future releases. Our immediate priorities include supporting periodic boundary
conditions and other basis sets such as grids, and plane waves.}
Overall, the TREXIO format and library is a valuable resource for the quantum chemistry community and its continued development and adoption will surely benefit the field.

\begin{acknowledgments}
\cf{
The authors would like to thank Susi Lehtola for providing valuable feedback on
an earlier version of this manuscript.}
 This work was supported by the European Centre of
Excellence in Exascale Computing TREX --- Targeting Real Chemical
Accuracy at the Exascale. \cf{Hence, the name of the software
is \emph{TREX Input/Output} (TREXIO).}
This project has received funding from the
European Union's Horizon 2020 --- Research and Innovation program ---
under grant agreement no.~952165.
A CC-BY 4.0 (\url{https://creativecommons.org/licenses/by/4.0/})
public copyright license has been applied by the authors to the present
document and will be applied to all subsequent versions up to the Author
Accepted Manuscript arising from this submission, in accordance with the
grant’s open access conditions.
\end{acknowledgments}

\bibliography{trexio}

\newpage
\appendix
\section{Table of stored data}

\begingroup
\squeezetable
\begin{longtable*}{llll}
  \caption{\label{tab:all}
List of all the data that can be stored in TREXIO files. The name of
the group is written in the first line of each block.
Multi-dimensional arrays are expressed in column-major order, meaning that elements of the same column are stored contiguously.
} \\

Attribute & Type & Dimensions & Description\\
\hline
\texttt{metadata} &  &  & \\
\texttt{code\_num} & \texttt{dim} &  & Number of codes used to produce the file\\
\texttt{code} & \texttt{str} & \texttt{(metadata.code\_num)} & Names of the codes used\\
\texttt{author\_num} & \texttt{dim} &  & Number of authors of the file\\
\texttt{author} & \texttt{str} & \texttt{(metadata.author\_num)} & Names of the authors of the file\\
\texttt{package\_version} & \texttt{str} &  & TREXIO version used to produce the file\\
\texttt{description} & \texttt{str} &  & Text describing the content of file\\
\texttt{unsafe} & \texttt{int} &  & \texttt{1}: true, \texttt{0}: false\\
\hline
\texttt{nucleus} &  &  & \\
\texttt{num} & \texttt{dim} &  & Number of nuclei\\
\texttt{charge} & \texttt{float} & \texttt{(nucleus.num)} & Charges of the nuclei\\
\texttt{coord} & \texttt{float} & \texttt{(3,nucleus.num)} & Coordinates of the atoms\\
\texttt{label} & \texttt{str} & \texttt{(nucleus.num)} & Atom labels\\
\texttt{point\_group} & \texttt{str} &  & Symmetry point group\\
\texttt{repulsion} & \texttt{float} &  & Nuclear repulsion energy\\
\hline
\texttt{grid} &  &  & \\
\texttt{description} & \texttt{str} &  & Details about the used quadratures can go here\\
\texttt{rad\_precision} & \texttt{float} &  & Radial precision parameter\\
\texttt{num} & \texttt{dim} &  & Number of grid points\\
\texttt{max\_ang\_num} & \texttt{int} &  & Maximum number of angular grid points\\
\texttt{min\_ang\_num} & \texttt{int} &  & Minimum number of angular grid points\\
\texttt{coord} & \texttt{float} & \texttt{(grid.num)} & Discretized coordinate space\\
\texttt{weight} & \texttt{float} & \texttt{(grid.num)} & Grid weights according to a given partitioning (e.g. Becke)\\
\texttt{ang\_num} & \texttt{dim} &  & Number of angular integration points\\
\texttt{ang\_coord} & \texttt{float} & \texttt{(grid.ang\_num)} & Discretized angular space\\
\texttt{ang\_weight} & \texttt{float} & \texttt{(grid.ang\_num)} & Angular grid weights\\
\texttt{rad\_num} & \texttt{dim} &  & Number of radial integration points\\
\texttt{rad\_coord} & \texttt{float} & \texttt{(grid.rad\_num)} & Discretized radial space\\
\texttt{rad\_weight} & \texttt{float} & \texttt{(grid.rad\_num)} & Radial grid weights\\
\hline
\texttt{electron} &  &  & \\
\texttt{num} & \texttt{dim} &  & Number of electrons\\
\texttt{up\_num} & \texttt{int} &  & Number of \(\uparrow\)-spin electrons\\
\texttt{dn\_num} & \texttt{int} &  & Number of \(\downarrow\)-spin electrons\\
\hline
\texttt{state} &  &  & \\
\texttt{num} & \texttt{dim} &  & Number of states (including the ground state)\\
\texttt{id} & \texttt{int} &  & Index of the current state (0 is ground state)\\
\texttt{current\_label} & \texttt{str} &  & Label of the current state\\
\texttt{label} & \texttt{str} & \texttt{(state.num)} & Labels of all states\\
\texttt{file\_name} & \texttt{str} & \texttt{(state.num)} & Names of the TREXIO files linked to the current one\\
\hline
\texttt{basis} &  &  & \\
\texttt{type} & \texttt{str} &  & Type of basis set: ``Gaussian'' or ``Slater''\\
\texttt{prim\_num} & \texttt{dim} &  & Total number of primitives\\
\texttt{shell\_num} & \texttt{dim} &  & Total number of shells\\
\texttt{nucleus\_index} & \texttt{index} & \texttt{(basis.shell\_num)} & One-to-one correspondence between shells and atomic indices\\
\texttt{shell\_ang\_mom} & \texttt{int} & \texttt{(basis.shell\_num)} & One-to-one correspondence between shells and angular momenta\\
\texttt{shell\_factor} & \texttt{float} & \texttt{(basis.shell\_num)} & Normalization factor of each shell (\(\mathcal{N}_s\))\\
\texttt{r\_power} & \texttt{int} & \texttt{(basis.shell\_num)} & Power to which \(r\) is raised (\(n_s\))\\
\texttt{shell\_index} & \texttt{index} & \texttt{(basis.prim\_num)} & One-to-one correspondence between primitives and shell index\\
\texttt{exponent} & \texttt{float} & \texttt{(basis.prim\_num)} & Exponents of the primitives (\(\gamma_{ks}\))\\
\texttt{coefficient} & \texttt{float} & \texttt{(basis.prim\_num)} & Coefficients of the primitives (\(a_{ks}\))\\
\texttt{prim\_factor} & \texttt{float} & \texttt{(basis.prim\_num)} & Normalization coefficients for the primitives (\(f_{ks}\))\\
\texttt{e\_cut} & \texttt{float} &  & Energy cut-off for plane-wave calculations\\
\hline
\texttt{ecp} &  &  & \\
\texttt{max\_ang\_mom\_plus\_1} & \texttt{int} & \texttt{(nucleus.num)} & \(\ell_{\max}+1\), in the removed core orbitals\\
\texttt{z\_core} & \texttt{int} & \texttt{(nucleus.num)} & Number of core electrons to remove per atom\\
\texttt{num} & \texttt{dim} &  & Total number of ECP functions for all atoms and all values of \(\ell\)\\
\texttt{ang\_mom} & \texttt{int} & \texttt{(ecp.num)} & One-to-one correspondence between ECP items and \(\ell\)\\
\texttt{nucleus\_index} & \texttt{index} & \texttt{(ecp.num)} & One-to-one correspondence between ECP items and the atom index\\
\texttt{exponent} & \texttt{float} & \texttt{(ecp.num)} & \(\alpha_{A q \ell}\) all ECP exponents\\
\texttt{coefficient} & \texttt{float} & \texttt{(ecp.num)} & \(\beta_{A q \ell}\) all ECP coefficients\\
\texttt{power} & \texttt{int} & \texttt{(ecp.num)} & \(n_{A q \ell}\) all ECP powers\\
\hline
\texttt{ao} &  &  & \\
\texttt{cartesian} & \texttt{int} &  & \texttt{1}: true, \texttt{0}: false\\
\texttt{num} & \texttt{dim} &  & Total number of atomic orbitals\\
\texttt{shell} & \texttt{index} & \texttt{(ao.num)} & Basis set shell for each AO\\
\texttt{normalization} & \texttt{float} & \texttt{(ao.num)} & Normalization factors $N^\prime$ \\
\hline
\texttt{ao\_1e\_int} &  &  & \\
\texttt{overlap} & \texttt{float} & \texttt{(ao.num,ao.num)} & \(\langle p \vert q \rangle\)\\
\texttt{kinetic} & \texttt{float} & \texttt{(ao.num,ao.num)} & \(\langle p \vert \hat{T}_{\text{e}} \vert q \rangle\)\\
\texttt{potential\_n\_e} & \texttt{float} & \texttt{(ao.num,ao.num)} & \(\langle p \vert \hat{V}_{\text{ne}} \vert q \rangle\)\\
\texttt{ecp} & \texttt{float} & \texttt{(ao.num,ao.num)} & \(\langle p \vert \hat{V}_{\text{ecp}} \vert q \rangle\)\\
\texttt{core\_hamiltonian} & \texttt{float} & \texttt{(ao.num,ao.num)} & \(\langle p \vert \hat{h} \vert q \rangle\)\\
\texttt{overlap\_im} & \texttt{float} & \texttt{(ao.num,ao.num)} & \(\langle p \vert q \rangle\) (imaginary part) \\
\texttt{kinetic\_im} & \texttt{float} & \texttt{(ao.num,ao.num)} & \(\langle p \vert \hat{T}_{\text{e}} \vert q \rangle\)   (imaginary part)\\
\texttt{potential\_n\_e\_im} & \texttt{float} & \texttt{(ao.num,ao.num)} & \(\langle p \vert \hat{V}_{\text{ne}} \vert q \rangle\)  (imaginary part)\\
\texttt{ecp\_im} & \texttt{float} & \texttt{(ao.num,ao.num)} & \(\langle p \vert \hat{V}_{\text{ECP}} \vert q \rangle\)  (imaginary part)\\
\texttt{core\_hamiltonian\_im} & \texttt{float} & \texttt{(ao.num,ao.num)} & \(\langle p \vert \hat{h} \vert q \rangle\) (imaginary part)\\
\hline
\texttt{ao\_2e\_int} &  &  & \\
\texttt{eri} & \texttt{float sparse} & \texttt{(ao.num,ao.num,ao.num,ao.num)} & Electron repulsion integrals\\
\texttt{eri\_lr} & \texttt{float sparse} & \texttt{(ao.num,ao.num,ao.num,ao.num)} & Long-range electron repulsion integrals\\
\texttt{eri\_cholesky\_num} & \texttt{dim} &  & Number of Cholesky vectors for ERI\\
\texttt{eri\_cholesky} & \texttt{float sparse} & \texttt{(ao.num,ao.num,ao\_2e\_int.eri\_cholesky\_num)} & Cholesky decomposition of the ERI\\
\texttt{eri\_lr\_cholesky\_num} & \texttt{dim} &  & Number of Cholesky vectors for long range ERI\\
\texttt{eri\_lr\_cholesky} & \texttt{float sparse} & \texttt{(ao.num,ao.num,ao\_2e\_int.eri\_lr\_cholesky\_num)} & Cholesky decomposition of the long range ERI\\
\hline
\texttt{mo} &  &  & \\
\texttt{type} & \texttt{str} &  & String to identify the set of MOs (HF, Natural, Local, CASSCF, \emph{etc})\\
\texttt{num} & \texttt{dim} &  & Number of MOs\\
\texttt{coefficient} & \texttt{float} & \texttt{(ao.num,mo.num)} & MO coefficients\\
\texttt{coefficient\_im} & \texttt{float} & \texttt{(ao.num,mo.num)} & MO coefficients (imaginary part)\\
\texttt{class} & \texttt{str} & \texttt{(mo.num)} & Choose among: Core, Inactive, Active, Virtual, Deleted\\
\texttt{symmetry} & \texttt{str} & \texttt{(mo.num)} & Symmetry in the point group\\
\texttt{occupation} & \texttt{float} & \texttt{(mo.num)} & Occupation number\\
\texttt{energy} & \texttt{float} & \texttt{(mo.num)} & For canonical MOs, corresponding eigenvalue\\
\texttt{spin} & \texttt{int} & \texttt{(mo.num)} & For UHF wave functions, 0 is \(\alpha\) and 1 is \(\beta\)\\
\hline
\texttt{mo\_1e\_int} &  &  & \\
\texttt{overlap} & \texttt{float} & \texttt{(mo.num,mo.num)} & \(\langle i \vert j \rangle\)\\
\texttt{kinetic} & \texttt{float} & \texttt{(mo.num,mo.num)} & \(\langle i \vert \hat{T}_e \vert j \rangle\)\\
\texttt{potential\_n\_e} & \texttt{float} & \texttt{(mo.num,mo.num)} & \(\langle i \vert \hat{V}_{\text{ne}} \vert j \rangle\)\\
\texttt{ecp} & \texttt{float} & \texttt{(mo.num,mo.num)} & \(\langle i \vert \hat{V}_{\text{ECP}} \vert j \rangle\)\\
\texttt{core\_hamiltonian} & \texttt{float} & \texttt{(mo.num,mo.num)} & \(\langle i \vert \hat{h} \vert j \rangle\)\\
\texttt{overlap\_im} & \texttt{float} & \texttt{(mo.num,mo.num)} & \(\langle i \vert j \rangle\) (imaginary part)\\
\texttt{kinetic\_im} & \texttt{float} & \texttt{(mo.num,mo.num)} & \(\langle i \vert \hat{T}_e \vert j \rangle\)   (imaginary part)\\
\texttt{potential\_n\_e\_im} & \texttt{float} & \texttt{(mo.num,mo.num)} & \(\langle i \vert \hat{V}_{\text{ne}} \vert j \rangle\)  (imaginary part)\\
\texttt{ecp\_im} & \texttt{float} & \texttt{(mo.num,mo.num)} & \(\langle i \vert \hat{V}_{\text{ECP}} \vert j \rangle\)  (imaginary part)\\
\texttt{core\_hamiltonian\_im} & \texttt{float} & \texttt{(mo.num,mo.num)} & \(\langle i \vert \hat{h} \vert j \rangle\) (imaginary part)\\
\hline
\texttt{mo\_2e\_ints} &  &  & \\
\texttt{eri} & \texttt{float sparse} & \texttt{(mo.num,mo.num,mo.num,mo.num)} & Electron repulsion integrals\\
\texttt{eri\_lr} & \texttt{float sparse} & \texttt{(mo.num,mo.num,mo.num,mo.num)} & Long-range electron repulsion integrals\\
\texttt{eri\_cholesky\_num} & \texttt{dim} &  & Number of Cholesky vectors for ERI\\
\texttt{eri\_cholesky} & \texttt{float sparse} & \texttt{(mo.num,mo.num,mo\_2e\_int.eri\_cholesky\_num)} & Cholesky decomposition of the ERI\\
\texttt{eri\_lr\_cholesky\_num} & \texttt{dim} &  & Number of Cholesky vectors for long range ERI\\
\texttt{eri\_lr\_cholesky} & \texttt{float sparse} & \texttt{(mo.num,mo.num,mo\_2e\_int.eri\_lr\_cholesky\_num)} & Cholesky decomposition of the long range ERI\\
\hline
\texttt{determinant} &  &  & \\
\texttt{num} & \texttt{dim readonly} &  & Number of determinants\\
\texttt{list} & \texttt{int special} & \texttt{(determinant.num)} & List of determinants as integer bit fields\\
\texttt{coefficient} & \texttt{float buffered} & \texttt{(determinant.num)} & Coefficients of the determinants from the CI expansion\\
\hline
\texttt{csf} &  &  & \\
\texttt{num} & \texttt{dim readonly} &  & Number of CSFs\\
\texttt{coefficient} & \texttt{float buffered} & \texttt{(csf.num)} & Coefficients \(C_I\) of the CSF expansion\\
\texttt{det\_coefficient} & \texttt{float sparse} & \texttt{(determinant.num,csf.num)} & Projection on the determinant basis\\
\hline
\texttt{amplitude} &  &  & \\
\texttt{single} & \texttt{float sparse} & \texttt{(mo.num,mo.num)} & Single excitation amplitudes\\
\texttt{single\_exp} & \texttt{float sparse} & \texttt{(mo.num,mo.num)} & Exponentialized single excitation amplitudes\\
\texttt{double} & \texttt{float sparse} & \texttt{(mo.num,mo.num,mo.num,mo.num)} & Double excitation amplitudes\\
\texttt{double\_exp} & \texttt{float sparse} & \texttt{(mo.num,mo.num,mo.num,mo.num)} & Exponentialized double excitation amplitudes\\
\texttt{triple} & \texttt{float sparse} & \texttt{(mo.num,mo.num,mo.num,mo.num,mo.num,mo.num)} & Triple excitation amplitudes\\
\texttt{triple\_exp} & \texttt{float sparse} & \texttt{(mo.num,mo.num,mo.num,mo.num,mo.num,mo.num)} & Exponentialized triple excitation amplitudes\\
\texttt{quadruple} & \texttt{float sparse} & \texttt{(mo.num,mo.num,mo.num,mo.num} & Quadruple excitation amplitudes\\
 &  & \texttt{,mo.num,mo.num,mo.num,mo.num)} & \\
\texttt{quadruple\_exp} & \texttt{float sparse} & \texttt{(mo.num,mo.num,mo.num,mo.num} & Exponentialized quadruple excitation amplitudes\\
 &  & \texttt{,mo.num,mo.num,mo.num,mo.num)} & \\
\hline
\texttt{rdm} &  &  & \\
\texttt{1e} & \texttt{float} & \texttt{(mo.num,mo.num)} & One body density matrix\\
\texttt{1e\_up} & \texttt{float} & \texttt{(mo.num,mo.num)} & \(\uparrow\)-spin component of the one body density matrix\\
\texttt{1e\_dn} & \texttt{float} & \texttt{(mo.num,mo.num)} & \(\downarrow\)-spin component of the one body density matrix\\
\texttt{2e} & \texttt{float sparse} & \texttt{(mo.num,mo.num,mo.num,mo.num)} & Two-body reduced density matrix (spin trace)\\
\texttt{2e\_upup} & \texttt{float sparse} & \texttt{(mo.num,mo.num,mo.num,mo.num)} & \(\uparrow \uparrow\) component of the two-body reduced density matrix\\
\texttt{2e\_dndn} & \texttt{float sparse} & \texttt{(mo.num,mo.num,mo.num,mo.num)} & \(\downarrow \downarrow\) component of the two-body reduced density matrix\\
\texttt{2e\_updn} & \texttt{float sparse} & \texttt{(mo.num,mo.num,mo.num,mo.num)} & \(\uparrow \downarrow\) component of the two-body reduced density matrix\\
\texttt{2e\_cholesky\_num} & \texttt{dim} &  & Number of Cholesky vectors\\
\texttt{2e\_cholesky} & \texttt{float sparse} & \texttt{(mo.num,mo.num,rdm.2e\_cholesky\_num)} & Cholesky decomposition of the two-body RDM (spin trace)\\
\texttt{2e\_upup\_cholesky\_num} & \texttt{dim} &  & Number of Cholesky vectors \(\uparrow \uparrow\)\\
\texttt{2e\_upup\_cholesky} & \texttt{float sparse} & \texttt{(mo.num,mo.num,rdm.2e\_upup\_cholesky\_num)} & Cholesky decomposition of the two-body RDM (\(\uparrow \uparrow\))\\
\texttt{2e\_dndn\_cholesky\_num} & \texttt{dim} &  & Number of Cholesky vectors \(\downarrow \downarrow\)\\
\texttt{2e\_dndn\_cholesky} & \texttt{float sparse} & \texttt{(mo.num,mo.num,rdm.2e\_dndn\_cholesky\_num)} & Cholesky decomposition of the two-body RDM (\(\downarrow \downarrow\))\\
\texttt{2e\_updn\_cholesky\_num} & \texttt{dim} &  & Number of Cholesky vectors \(\uparrow \downarrow\)\\
\texttt{2e\_updn\_cholesky} & \texttt{float sparse} & \texttt{(mo.num,mo.num,rdm.2e\_updn\_cholesky\_num)} & Cholesky decomposition of the two-body RDM (\(\uparrow \downarrow\))\\
\hline
\texttt{jastrow} &  &  & \\
\texttt{type} & \texttt{string} &  & Type of Jastrow factor: \texttt{CHAMP} or \texttt{Mu}\\
\texttt{ee\_num} & \texttt{dim} &  & Number of electron-electron parameters\\
\texttt{en\_num} & \texttt{dim} & & Number of electron-nucleus parameters, per nucleus\\
\texttt{een\_num} & \texttt{dim} & & Number of electron-electron-nucleus parameters, per nucleus\\
\texttt{ee} & \texttt{float} & \texttt{(jastrow.ee\_num)} & Electron-electron parameters\\
\texttt{en} & \texttt{float} & \texttt{(jastrow.en\_num)} & Electron-nucleus parameters\\
\texttt{een} & \texttt{float} & \texttt{(jastrow.een\_num)} & Electron-electron-nucleus parameters\\
\texttt{en\_nucleus} & \texttt{index} & \texttt{(jastrow.en\_num)} & Nucleus relative to the eN parameter\\
\texttt{een\_nucleus} & \texttt{index} & \texttt{(jastrow.een\_num)} & Nucleus relative to the eeN parameter\\
\texttt{ee\_scaling} & \texttt{float} &  & \(\kappa\) value in CHAMP Jastrow for electron-electron distances\\
\texttt{en\_scaling} & \texttt{float} & \texttt{(nucleus.num)} & \(\kappa\) value in CHAMP Jastrow for electron-nucleus distances\\
\hline
\texttt{qmc} &  &  & \\
\texttt{num} & \texttt{dim} &  & Number of 3N-dimensional points\\
\texttt{point} & \texttt{float} & \texttt{(3,electron.num,qmc.num)} & 3N-dimensional points\\
\texttt{psi} & \texttt{float} & \texttt{(qmc.num)} & Wave function evaluated at the points\\
\texttt{e\_loc} & \texttt{float} & \texttt{(qmc.num)} & Local energy evaluated at the points

\end{longtable*}
\endgroup

\end{document}